\documentclass[graybox]{svmult}

\usepackage{mathptmx}       
\usepackage{helvet}         
\usepackage{courier}        
\usepackage{type1cm}        
\usepackage{makeidx}         
\usepackage{graphicx}        
\usepackage{multicol}        
\usepackage[bottom]{footmisc}
\usepackage{tabularx}
\usepackage{booktabs}
\usepackage{pdflscape}
\usepackage{url}
\usepackage{float}
\usepackage[numbers,sort]{natbib} 

\makeindex             

\newcolumntype{Y}{>{\centering\arraybackslash}X}

\begin{document}

\title*{Incorporating Emotion and Personality-Based Analysis in User-Centered Modelling}
\author{Mohamed Mostafa \and Tom Crick \and Ana C. Calderon \and Giles Oatley}
\institute{Mohamed Mostafa \and Tom Crick \and Ana C. Calderon \at
  Department of Computing \& Information Systems, Cardiff Metropolitan
  University, Cardiff, UK;
  \email{{momostafa,tcrick,acalderon}@cardiffmet.ac.uk}
\and
Giles Oatley \at School of Engineering \& Information Technology,
Murdoch University, Australia;\\\email{g.oatley@murdoch.edu.au}}
%
%
\maketitle

\abstract{Understanding complex user behaviour under various
conditions, scenarios and journeys can be fundamental to the
improvement of the user-experience for a given system. Predictive
models of user reactions, responses -- and in particular, emotions --
can aid in the design of more intuitive and usable systems. Building
on this theme, the preliminary research presented in this paper
correlates events and interactions in an online social network against
user behaviour, focusing on personality traits. Emotional context and
tone is analysed and modelled based on varying types of sentiments
that users express in their language using the IBM Watson Developer
Cloud tools. The data collected in this study thus provides further
evidence towards supporting the hypothesis that analysing and
modelling emotions, sentiments and personality traits provides
valuable insight into improving the user experience of complex social
computer systems.}


\section{Introduction}\label{intro}

As computer systems and applications have become more widespread and
complex, with increasing demands and expectations of ever-more
intuitive human-computer interactions, research in modelling,
understanding and predicting user behaviour demands has become a
priority across a number of domains.  In these application domains, it
is useful to obtain knowledge about user profiles or models of
software applications, including intelligent agents, adaptive systems,
intelligent tutoring systems, recommender systems, e-commerce
applications and knowledge management
systems~\cite{schiaffino+amandi:2009}. Furthermore, understanding user
behaviour during system events leads to a better informed predictive
model capability, allowing the construction of more intuitive
interfaces and an improved user experience. This work can be applied
across a range of socio-technical systems, impacting upon both
personal and business computing.

We are particularly interested in the relationship between digital
footprint and behaviour and
personality~\citep{oatley+crick:2014,oatley-et-al_dasc2015,blamey-et-al-2012,blamey-et-al-2013}.
A wide range of pervasive and often publicly available datasets
encompassing digital footprints, such as social media activity, can be
used to infer
personality~\citep{lambiotte+kosinski:2014,oatley-et-al-soccogcomp2015}.
Big social data offers the potential for new insights into human
behaviour and development of robust models capable of describing
individuals and societies~\citep{lazer-et-al:2009}. Social media has
been used in varying computer system approaches; in the past this has
mainly been the textual information contained in blogs, status posts
and photo comments~\cite{blamey-et-al-2012,blamey-et-al-2013}, but
there is also a wealth of information in the other ways of interacting
with online artefacts. Research in image or video analysis includes
promising studies on YouTube videos for classification of specific
behaviours and indicators of personality
traits~\citep{biel+gatica-perez:2012}. This work uses crowdsourced
impressions, social attention, and audiovisual behavioural analysis on
slices of conversational video blogs extracted from YouTube. From
sharing and gathering of information and data, to catering for
marketing and business needs; it is now widely used as technical
support for computer system platforms.



The work presented in this paper is building upon previous work in
psycholinguistic science (the study of the psychological and
neurobiological factors that enable humans to acquire, use, comprehend
and produce language) and aims to provide further insight into how the
words and constructs we use in our daily life and online interactions
reflect our personalities and our underlying emotions. As part of this
active research field, it is widely accepted that written text
reflects more than the words and syntactic constructs, but also
conveys emotion and personality
traits~\citep{pennebaker+king:1999}. As part of our work, the IBM
Watson Tone Analyzer (part of the IBM Watson Developer Cloud
toolchain) has been used to identify emotion tones in the textual
interactions in an online system, building on previous work in this
area that shows a strong correlation between the word choice and
personality, emotions, attitude and cognitive processes, providing
further evidence that it is possible to profile and potentially
predict users’ identity~\citep{fast+funder:2008}. The
{\emph{Linguistic Inquiry and Word Count}} (LIWC) psycholinguistics
dictionary~\citep{pennebaker-et-al:2001,tausczik+pennebaker:2010} is
used to find psychologically meaningful word categories from word
usage in writing; the work presented here provides a modelling and
analysis framework, as well as associated toolchain, for further
application to larger datasets to support the research goal of
improving user-centered modelling.

The paper is structured as follows: in Section~\ref{personality} we
give an overview of how insight into personality can be inferred from
textual data; in Section~\ref{tools} we introduce the tools, namely
the IBM Watson Developer Cloud toolchain; in
Section~\ref{dataanalysis} we present our data and the statistical
analysis; in Section~\ref{model} we identify the key elements of our
model and in Sections~\ref{conclusions} and \ref{futurework} we
highlight the conclusions and main contributions of this paper, as
well as clear recommendations for future research.

\section{Personality Insight}\label{personality}

Numerous studies have suggested key words and phrases can signal
underlying tendencies and that this can form the basis of identifying
certain aspects of
personality~\citep{iacobelli-et-al:2011,pennebaker+king:1999,oberlander+gill:2004,oberlander+gill:2006}.
\citet{scherer:1984} introduced a valuable classification with the
following distinctions between emotions, moods, interpersonal stances,
attitudes and personality traits:

\begin{itemize}
\item {\emph{Emotion:}} short-lived, for instance being angry, sad, or joyful;
\item {\emph{Mood:}} longer-lived, low-intensity for instance being cheerful or gloomy;
\item {\emph{Interpersonal stances:}} duration linked to specific
  interaction, for instance friendly or supportive;
\item {\emph{Attitudes:}} long-lived linked to specific people or
  events, for instance loving and hating;
\item {\emph{Personality traits:}} stable personality dispositions and
  typical behaviour tendencies, for instance nervous, anxious, or hostile.
\end{itemize}

By observing the occurrences of words that relate to these five
categories, we can conclude infer aspects of the holder's
psychological state. For instance, we have sentiment analysis or
opinion mining, utilising open-source software such as
{\emph{SentiWordNet}}; as well as the use of features from
psycholinguistic databases such as
{\emph{LIWC}}~\citep{pennebaker-et-al:2001} to create a range of
statistical models for each of the Five Factor personality
traits~\citep{mairesse-et-al:2007}. This ``Big Five'' model, focuses
on five dimensions, namely: {\emph{Agreeableness}},
{\emph{Conscientiousness}}, {\emph{Extraversion}},
{\emph{Neuroticism}} and
{\emph{Openness}}~\citep{norman:1963}: 

\begin{itemize}
\item {\emph{Agreeableness}} (friendly/compassionate
  vs. analytical/detached) is the tendency to be compassionate and
  cooperative rather than suspicious and antagonistic. It is also a
  measure of one's trusting and helpful nature; high agreeableness is
  often seen as naive or submissive, whereas low agreeableness
  personalities are often competitive or challenging people, which can
  be seen as argumentative or untrustworthy.
\item {\emph{Conscientiousness}} (efficient/organized
  vs. easy-going/careless) is the tendency to be organised and
  dependable, showing self-discipline, acting dutifully, aiming for
  achievement, and preferring planned rather than spontaneous
  behaviour. High conscientiousness is often perceived as stubborn and
  obsessive, whereas low conscientiousness personalities are flexible
  and spontaneous, but can be perceived as sloppy and unreliable.
\item {\emph{Extraversion}} (outgoing/energetic vs. solitary/reserved)
  is the tendency to seek stimulation in the company of others, and
  talkativeness; high extraversion is often perceived as
  attention-seeking and domineering, whereas low extraversion causes a
  reserved, reflective personality, which can be perceived as aloof or
  self-absorbed.
\item {\emph{Neuroticism}} (sensitive/nervous vs. secure/confident) is
  the tendency to experience unpleasant emotions easily, such as
  anger, anxiety, depression, and vulnerability. Neuroticism also
  refers to the degree of emotional stability and impulse control and
  is sometimes referred to as ``emotional stability''. 
\item {\emph{Openness to experience}} (inventive/curious
  vs. consistent/cautious) reflects the degree of intellectual
  curiosity, creativity and a preference for novelty and variety a
  person has. High openness can be perceived as unpredictability or
  lack of focus, whereas those with low openness seek to gain
  fulfillment through perseverance, and are characterized as pragmatic
  and data-driven—sometimes even perceived to be dogmatic and
  closed-minded. Some disagreement remains about how to interpret and
  contextualise the openness factor~\citep{peabody+goldberg:1989}.
\end{itemize}

It should be noted that while researchers have continued to work with
the Five Factors model, there are well known
limitations~\cite{eysenck:1992,paunonen+jackson:2000,block:2010} that
are often overlooked; however, over the past fifty years the Five
Factor model has become a standard in
psychology~\cite{mairesse-et-al:2007}, developing a large corpus of
work to compare against.

\section{IBM Watson Developer Cloud Tools}\label{tools}

The IBM Watson Developer
Cloud\footnote{\url{http://www.ibm.com/watson/developercloud/}} offers
a variety of services across language, speech, vision and data insight
for developing cognitive applications; each Watson service provides a
REST API for interacting with the service. In this paper, we have used
two of the language services -- Tone Analyzer and Personality Insights
-- but also includes natural language classifiers, language
translation and retrieve and rank, a machine learning information
retrieval tool.

\subsection{IBM Watson Tone Analyzer}

The IBM Watson Tone
Analyzer\footnote{\url{http://www.ibm.com/watson/developercloud/tone-analyzer.html}}
is a cloud-based framework to infer emotions from a given text; its
use cases include: personal and business communications, market
research, self-branding and automated contact-center agents.  It uses
linguistic analysis to detect three types of tones from written text:
emotions, social tendencies, and writing style. Emotions identified
include {\emph{Anger}}, {\emph{Fear}}, {\emph{Joy}}, {\emph{Sadness}}
and {\emph{Disgust}}; identified social tendencies include the Big
Five personality traits (as described above); identified writing
styles include {\emph{Confident}}, {\emph{Analytical}} and
{\emph{Tentative}}.

To derive emotion scores from text, IBM Watson Tone Analyzer uses a
stacked generalisation-based ensemble framework to achieve greater
predictive accuracy~\citep{costa+mccrae:1992}.  Features such as
n-grams (unigrams, bigrams and trigrams), punctuation, emoticons,
curse words, greeting words (such as ``{\emph{hello}}'',
``{\emph{hi}}'' and ``{\emph{thanks}}'') and sentiment polarity are
fed into machine learning algorithms to classify emotion
categories~\citep{fellbaum:2006}. The analysis performed by the Tone
Analyzer service is different from sentiment and emotion
analyses. Sentiment analysis can identify the positive and negative
sentiments within a document or web page. The sentiments can include
document-level sentiment, sentiment for a user-specified target,
entity-level sentiment, and keyword-level sentiment. Emotion analysis
can infer different categories of emotions such as joy, anger,
disgust, sadness, and fear. The Tone Analyzer service computes
emotional tones, in addition to social and writing style tones. LIWC
is then used to find psychologically meaningful word categories from
word usage in writing.

\subsection{IBM Watson Personality Insights}

IBM Watson Personality
Insights\footnote{\url{http://www.ibm.com/watson/developercloud/personality-insights.html}}
provides a deeper understanding of people's personality
characteristics, needs, and values to drive personalisation. It
enables applications to derive insights from social media, enterprise
data, or other digital communications by extracting and analysing a
spectrum of personality attributes to help discover actionable
insights about people and entities. The service can automatically
infer, from potentially noisy social media, portraits of individuals
that reflect their personality characteristics.

The service outputs personality characteristics that are divided into
three dimensions: the Big Five, {\emph{Values}} and {\emph{Needs}}. As
described in Section~\ref{personality}, the five primary (top-level)
dimensions ({\emph{Agreeableness}}, {\emph{Conscientiousness}},
{\emph{Extraversion}}, {\emph{Emotional Range}} and {\emph{Openness}})
have six subdimensions, or facets, that further characterise an
individual according to the dimension. A Personality Insights portrait
can only be created where sufficient data of suitable quantity and
quality is provided. Because language use varies naturally from
document to document, a small sample of text might not be
representative of an individual's overall language patterns. Moreover,
different characteristics and different media converge at somewhat
different rates. While some services are contextually specific
depending on the domain model and content, Personality Insights
usually only requires a minimum of 3000+ words of any text, but
ideally reflective in nature. The percentile and sampling error
provide normalised scores that describe the extent to which the
author's writing exhibits a characteristic and the possible range of
deviation. They indicate an interval in which the service has 95\%
confidence in its results; the raw score and raw sampling error
provide similar results.


\section{Data Analysis \& Feature Extraction}\label{dataanalysis}

\subsection{Overview of the Data}

Our dataset comes from an online portal for a European Union 
international scholarship mobility hosted at a UK university. The
dataset was generated from interactions between users and a complex
online information system, namely the online portal for submitting
applications.

The whole dataset consists of users (n=391), interactions and comments
(n=1390) as responses to system status and reporting their experience
with using the system. Google Analytics has been used to track user behaviour
and web statistics (such as impressions); this data from has been used
to identify the server's status and categorised the status as two
stages: {\emph{Idle}, where the system had a higher number of active
sessions; and marked as {\emph{Failure}}, where the system had a lower
number of sessions engaged. Figure~\ref{fig:googleanalytics} provides
a plot of web traffic from Google Analytics over a specific day, clearly
showing the drop at 20:00 where the system had been identified as in
the {\emph{Failure}} state.

\begin{figure}[!ht]
\centering
\includegraphics[width=\columnwidth]{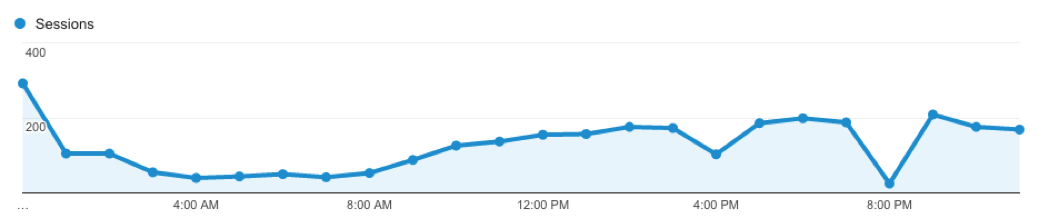}
\caption{Google Analytics profile shows behaviour of the system over a
  24 hour period (timeline during the day vs. number of active sessions)}
\label{fig:googleanalytics} 
\end{figure}

All interactions had been collected and grouped by server status, then
sent to the IBM Watson Tone Analyzer to generate the emotion social
tone scores, to provide an overview of the system behaviour and user’s
interaction with Facebook at the same
time. Figure~\ref{fig:emotiontone} shows the relationship between the
server behaviour and emotions of the users; in the system,
{\emph{Failure}} status shows a significant difference in overall
{\emph{Anger}} in different status; furthermore, the {\emph{Joy}}
parameter shows a significant difference with the system in
{\emph{Idle}} and {\emph{Failure}} status. However {\emph{Fear}} and
{\emph{Sadness}} parameters is almost the same, even with the system in
{\emph{Idle}} status.

\begin{figure}[!ht]
\centering
\includegraphics[width=\columnwidth]{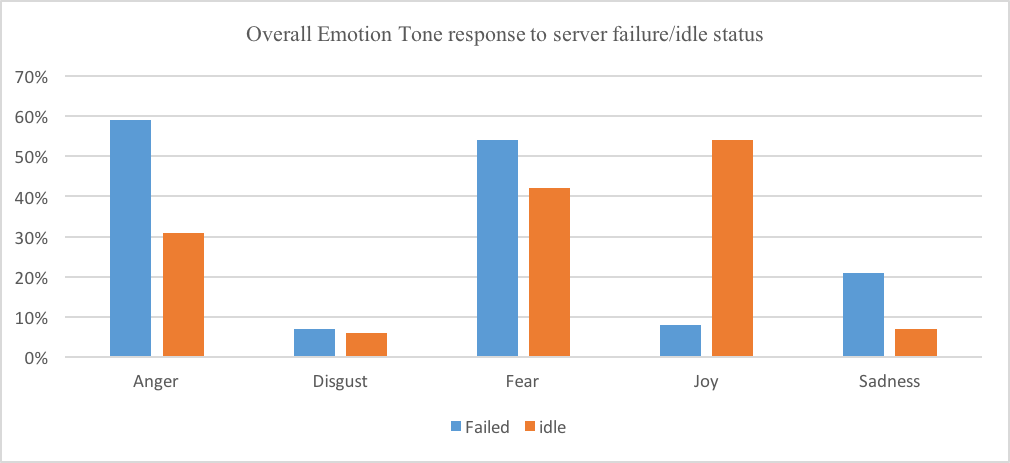}
\caption{Overall emotion tone response to server failure/idle status}
\label{fig:emotiontone} 
\end{figure}

We identified the user's personality based on analysis of their
Facebook interactions, namely by collecting all comments from the
users, again using the IBM Watson Personality Insights tool. However,
a number of users in the dataset had completed the Big Five
questionnaire (n=44); for these users, their Big Five scores have
been used instead. The second stage involved grouping the comments
based on server status and segmenting these interactions by user; this
allowed us to investigate the impact of server status in the emotion
of the user and investigate the Big Five dimension as a constant
parameter. By investigating the relationship between personality trait
dimensions and the social emotion tones, we are able to find the
highest correlation to identify the key elements of the potential
model by applying linear regression and Pearson correlation. This will
allow building of a neural network multilayer perception using the
potential key elements with higher correlations.

The previous overview encourages further investigation to understand
the relationship between user's behaviour and complex computer system
behaviours. The data collected from the social media interactions have
been grouped by users and using the IBM Watson Personality Insights,
we were able to identify the Big Five personality traits for each
user. Using the IBM Watson Tone Analyzer, the data has been grouped by
user's comments and server status ({\emph{Failure}}, {\emph{Idle}}) to
identify social emotion tone for each user. Table~\ref{tab:sample}
shows a sample of data used in this analysis, with each row
representing a unique user, and each column represents the Big Five
traits, social emotion tones and server status.

\begin{table}[ht]
\centering
\resizebox{\columnwidth}{!}{%
\begin{tabular}{@{}rrrrrrrrrrr@{}}
\toprule
\multicolumn{1}{c}{Openness} & \multicolumn{1}{c}{Conscientiousness} & \multicolumn{1}{c}{Extraversion} & \multicolumn{1}{c}{Agreeableness} & \multicolumn{1}{c}{Neuroticism} & \multicolumn{1}{c}{anger} & \multicolumn{1}{c}{disgust} & \multicolumn{1}{c}{fear} & \multicolumn{1}{c}{joy} & \multicolumn{1}{c}{sadness} & \multicolumn{1}{c}{Server  Status} \\ 
\midrule
0.528 & 0.523 & 0.537 & 0.653 & 0.511 & 0.217821 & 0.793375 & 0.501131 & 0.031477 & 0.284936 & Failure \\
0.252 & 0.063 & 0.037 & 0.266 & 0.989 & 0.542857 & 0.084615 & 0.178302 & 0.224453 & 0.264283 & Failure \\
0.817 & 0.571 & 0.157 & 0.012 & 0.401 & 0.162798 & 0.166694 & 0.213870 & 0.410916 & 0.220049 & Failure \\
0.197 & 0.130 & 0.180 & 0.419 & 0.990 & 0.468938 & 0.259794 & 0.350803 & 0.037265 & 0.636412 & Failure \\
0.155 & 0.079 & 0.081 & 0.226 & 0.975 & 0.539162 & 0.219993 & 0.431932 & 0.011625 & 0.642158 & Failure \\
0.158 & 0.281 & 0.332 & 0.510 & 0.869 & 0.419015 & 0.162022 & 0.213941 & 0.066892 & 0.686369 & Failure \\
0.817 & 0.571 & 0.157 & 0.012 & 0.401 & 0.041602 & 0.026298 & 0.141606 & 0.651962 & 0.106500 & Failure \\
0.058 & 0.038 & 0.147 & 0.375 & 0.989 & 0.449222 & 0.057946 & 0.181654 & 0.158412 & 0.547968 & Idle \\
0.178 & 0.138 & 0.800 & 0.564 & 0.828 & 0.207497 & 0.096643 & 0.093218 & 0.769316 & 0.162241 & Idle \\
0.105 & 0.463 & 0.792 & 0.704 & 0.041 & 0.134487 & 0.257145 & 0.195858 & 0.181699 & 0.509379 & Idle \\
0.589 & 0.479 & 0.147 & 0.339 & 0.828 & 0.360527 & 0.240875 & 0.321188 & 0.117492 & 0.212762 & Idle \\
0.338 & 0.235 & 0.104 & 0.304 & 0.869 & 0.164107 & 0.015058 & 0.230148 & 0.629562 & 0.356028 & Idle \\
0.204 & 0.203 & 0.480 & 0.329 & 0.892 & 0.625891 & 0.193692 & 0.242459 & 0.153679 & 0.166561 & Idle \\
0.689 & 0.968 & 0.805 & 0.465 & 0.029 & 0.246246 & 0.080353 & 0.123761 & 0.807537 & 0.135646 & Idle \\
0.093 & 0.175 & 0.642 & 0.563 & 0.875 & 0.279503 & 0.045658 & 0.207278 & 0.088724 & 0.505607 & Idle \\
0.277 & 0.296 & 0.276 & 0.332 & 0.892 & 0.499199 & 0.143897 & 0.269725 & 0.188664 & 0.285462 & Idle \\
0.055 & 0.095 & 0.783 & 0.699 & 0.935 & 0.450997 & 0.153940 & 0.263070 & 0.350778 & 0.116282 & Idle \\ 
\bottomrule
\end{tabular}%
}
\caption{Example data snapshot used in the analysis}
\label{tab:sample}
\end{table}

\subsection{Statistical Analysis}

As part of modelling the users' responses and behaviour, one of the
approaches to building a conceptual framework model is to apply linear
regression to investigate the relationship between the Big Five
personality dimensions and the emotion tones features.


\begin{figure}[H]
\centering
\includegraphics[width=0.8\columnwidth]{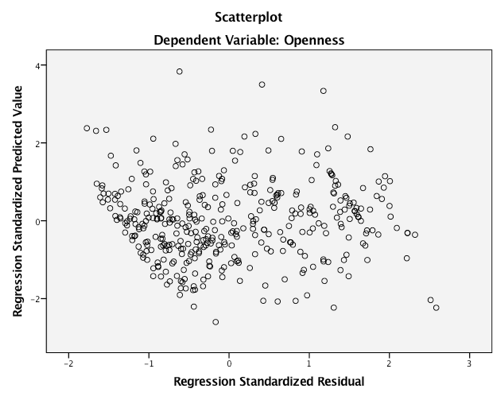}
\caption{Scatterplots of Big Five dimension ``Openness'' (dependent variable) and
  social emotion tones (independent variables)}
\label{fig:opennessplot} 
\end{figure}

\begin{figure}[H]
\centering
\includegraphics[width=0.8\columnwidth]{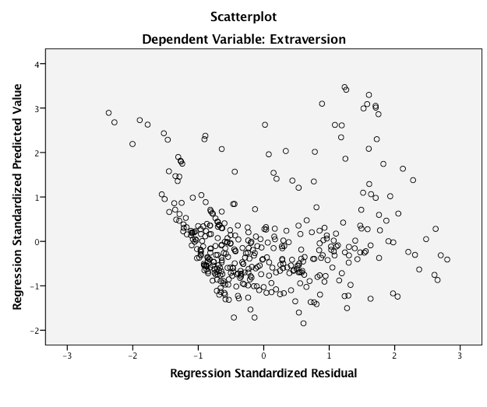}
\caption{Scatterplots of Big Five dimension ``Extraversion'' (dependent variable) and
  social emotion tones (independent variables)}
\label{fig:extraversionplot}
\end{figure}

\begin{figure}[H]
\centering
\includegraphics[width=0.8\columnwidth]{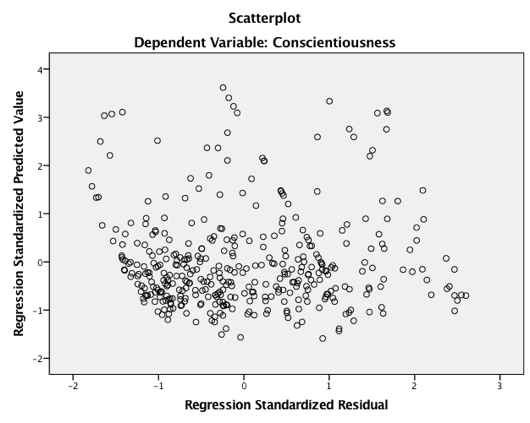}
\caption{Scatterplots of Big Five dimension ``Conscientiousness'' (dependent variable) and
  social emotion tones (independent variables)}
\label{fig:conscientiousnessplot} 
\end{figure}

\begin{figure}[H]
\centering
\includegraphics[width=0.8\columnwidth]{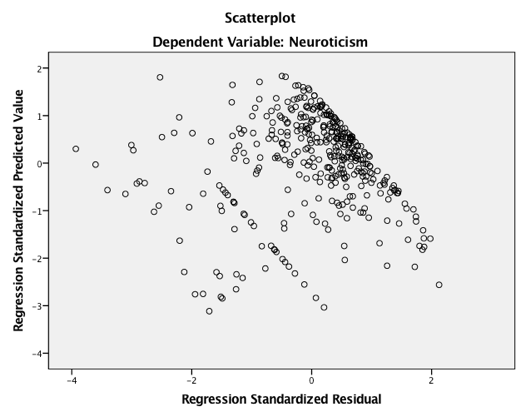}
\caption{Scatterplots of Big Five dimension ``Neuroticism'' (dependent variable) and
  social emotion tones (independent variables)}
\label{fig:neuroticismplot} 
\end{figure}

\begin{figure}[H]
\centering
\includegraphics[width=0.8\columnwidth]{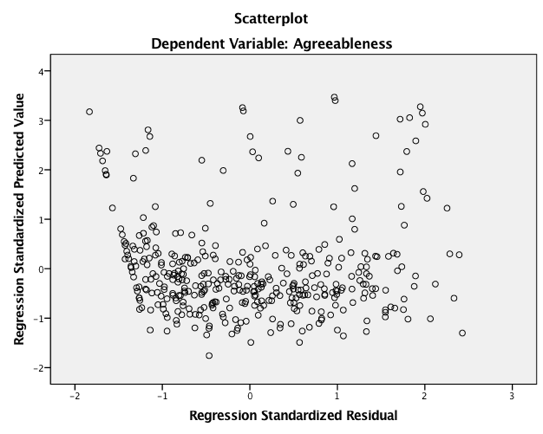}
\caption{Scatterplots of Big Five dimension ``Agreeableness'' (dependent variable) and
  social emotion tones (independent variables)}
\label{fig:agreeablenessplot} 
\end{figure}

\begin{table}[ht]
\centering
\resizebox{\columnwidth}{!}{%

\begin{tabular}{@{}llll|lll|lll|lll|lll@{}}
\toprule
& \multicolumn{3}{c}{Openness} & \multicolumn{3}{c}{Extraversion} &
\multicolumn{3}{c}{Conscientiousness} &
\multicolumn{3}{c}{Agreeableness} & \multicolumn{3}{c}{Neuroticism} \\

\midrule
           
& \multicolumn{1}{c}{B} & \multicolumn{1}{c}{t} &
\multicolumn{1}{c}{Sig} & \multicolumn{1}{c}{B} &
\multicolumn{1}{c}{t} & \multicolumn{1}{c}{Sig} &
\multicolumn{1}{c}{B} & \multicolumn{1}{c}{t} &
\multicolumn{1}{c}{Sig} & \multicolumn{1}{c}{B} &
\multicolumn{1}{c}{t} & \multicolumn{1}{c}{Sig} &
\multicolumn{1}{c}{B} & \multicolumn{1}{c}{t} &
\multicolumn{1}{c}{Sig} \\

(constant) & 0.356                 & 3.282                 & 0.001                   & 0.162                 & 1.642                 & 0.101                   & 0.16                  & 1.623                 & 0.105                   & 0.297      & 2.831     & 0.005    & 0.828          & 9.934   & 0      \\
anger      & -0.063                & -0.735                & 0.463                   & 0.064                 & 0.831                 & 0.406                   & 0.124                 & 1.592                 & 0.112                   & 0.024      & 0.293     & 0.769    & 0.116          & 1.767    & 0.078     \\
disgust    & 0.478                 & 4.354                 & 0                       & 0.114                 & 1.142                 & 0.253                   & 0.255                 & 2.551                 & 0.011                   & -0.061     & -0.574    & 0.566    & -0.363         & -4.303    & 0    \\
fear       & 0.065                 & 0.534                 & 0.594                   & 0.172                 & 1.549                 & 0.122                   & 0.04                  & 0.356                 & 0.722                   & 0.093      & 0.783     & 0.434    & -0.023         & -0.241    & 0.81    \\
joy        & 0.066                 & 0.561                 & 0.575                   & 0.446                 & 4.179                 & 0                       & 0.436                 & 4.058                 & 0                       & 0.188      & 1.652     & 0.099    & -0.487         & -5.39   & 0      \\
sadness    & -0.226                & -2.118                & 0.035                   & -0.185                & -1.906                & 0.057                   & -0.03                 & -0.313                & 0.754                   & 0.014      & 0.132     & 0.895    & 0.233          & 2.841     & 0.005    \\

\bottomrule

\end{tabular}%
}
\caption{Linear regression coefficients}
\label{tbl:linreg}
\end{table}

During the analysis, the linear regressions (presented in
Table~\ref{tbl:linreg} and Figures~\ref{fig:opennessplot},
\ref{fig:extraversionplot}, \ref{fig:conscientiousnessplot},
\ref{fig:neuroticismplot} and \ref{fig:agreeablenessplot}) do not show
significant correlations between the Big Five dimensions and the
social emotion tones; however, certain correlations can be highlighted
and used as key elements for the model at this stage. The correlation
of {\emph{Openness} and {\emph{Disgust}}, is 0.479; the correlation of
{\emph{Extraversion}} and {\emph{Joy}} is 0.446 with p-value of
zero. {\emph{Conscientiousness}} and {\emph{Joy}} with 0.436
correlation and \emph{Disgust}} with 0.255. {\emph{Agreeableness}},
does not appear to have a high impact in the social emotion
parameters, with the highest correlation being 0.188 with
{\emph{Joy}}, which can be overlooked as a useful factor in the
model. {\emph{Neuroticism}} and {\emph{Disgust} is -0.363, {\emph{Joy}
is -0.487 and p-value is zero is both cases; and {\emph{Sadness}} with
0.233. All correlation values are $<$0.5; however, it is noticed that
{\emph{Agreeableness}} does not have a linear relationship with any of
the social emotion tones. Furthermore, the social emotion tones that
have a potential linear relationship are {\emph{Disgust}},
{\emph{Joy}} and {\emph{Sadness}}, since the three tones have a
correlation between $>$0.3 and $<$0.5.

Previous linear regression analysis suggested that the following Big
Five dimensions ({\emph{Openness}}, {\emph{Extraversion}},
{\emph{Conscientiousness}} and {\emph{Neuroticism}}) have the highest
correlation with the social emotion tones ({\emph{Joy}},
{\emph{Sadness}} and {\emph{Disgust}}). For further analysis, the
Pearson correlation for the same dataset has been performed to compare
the output with the linear regression correlations. As you can see in
Table~\ref{tab:pearson}, there is no significant correlation in both;
however, in the Pearson correlation, {\emph{Neuroticism}} has the
highest correlation values across emotion tones, especially
{\emph{Anger}}, {\emph{Joy}} and {\emph{Sadness}}. {\emph{Joy}} does
have a correlation with all Big Five dimensions except for
{\emph{Agreeableness}} which agrees with the previous
analysis. However, {\emph{Disgust}} does not have a strong correlation
with any of the Big Five dimensions, which deviates from the previous
analysis.

\begin{table}[!ht]
\centering
\begin{tabular}{@{}rrrrrr@{}}
\toprule
                  & \multicolumn{1}{c}{Anger}  & \multicolumn{1}{c}{Disgust} & \multicolumn{1}{c}{Fear}   & \multicolumn{1}{c}{Joy}    & \multicolumn{1}{c}{Sadness} \\ 
\midrule
Openness          & -0.098 & 0.231   & 0.043  & 0.035  & -0.151  \\
Conscientiousness & -0.111 & -0.001  & -0.113 & 0.267  & -0.19   \\
Extraversion      & -0.175 & -0.077  & -0.071 & 0.349  & -0.291  \\
Agreeableness     & -0.068 & -0.089  & -0.027 & 0.14   & -0.069  \\
Neuroticism       & 0.375  & -0.037  & 0.153  & -0.488 & 0.379   \\ 

\bottomrule
\end{tabular}
\caption{Pearson correlations}
\label{tab:pearson}
\end{table}

\section{Key Elements of the Model}\label{model}

According to the output of the statistical analysis presented in
Table~\ref{tbl:linreg} (linear regression) and Table~\ref{tab:pearson}
(Pearson correlation), the Big Five dimension identified as the key
elements from the personality traits are: {\emph{Openness}},
{\emph{Extraversion}}, {\emph{Conscientiousness}} and
{\emph{Neuroticism}}. The statistical analysis agrees that
{\emph{Agreeableness}} does not have a significant correlation across
any of the social emotion tones. The social emotion tones to be used
as key input elements for the proposed model are {\emph{Joy}},
{\emph{Sadness}}, {\emph{Anger}} and {\emph{Disgust}}; although the
{\emph{Anger}} tone did not show any significant correlation in linear
regression analysis, the value of the Pearson correlation coefficient
is between 0.3 and 0.5 which can be used as input for the model.

\begin{table}[!ht]
\centering
\begin{tabular}{lrr}
Correctly classified instances:   & 43     & ({\emph{75.44\%}}) \\
Incorrectly classified instances: & 14     & ({\emph{24.56\%}}) \\
Kappa statistic:                  & 0.5295 &         \\
Mean absolute error:              & 0.3432 &         \\
Root mean squared error:          & 0.4246 &         \\
Total number of instances:        & 57     &        
\end{tabular}
\caption{Re-evaluation output of proposed model}
\label{tab:reeval}
\end{table}

The dataset used to build this model is based upon a number of users
(n=391), eight inputs ({\emph{Openness}}, {\emph{Extraversion}},
{\emph{Conscientiousness}}, {\emph{Neuroticism}}, {\emph{Joy}},
{\emph{Sadness}}, {\emph{Anger}} and {\emph{Disgust}}) and the
class/output variable as the server status (where No: System
{\emph{Failure}} and Yes: System {\emph{Idle}}). As shown in
Table~\ref{tab:reeval}, the total number of the instances for the
testing set is 57. The output of the model shows a 75.44\% corrected
predicted instances and 24.56\% incorrectly classified instances. As
this has been performed on a small subset of the overall larger
project dataset, the output data is encouraging and provides the
infrastructure for further analysis and research to exploit the full
dataset.

\section{Conclusions}\label{conclusions}

This paper presents preliminary results from a larger ongoing theme of
research to profile online/digital
behaviour~\citep{oatley+crick:2014,oatley+crick-gisruk2015,oatley-et-al_dasc2015},
which could provide the conceptual framework to improve user
experience and computer system architecture design. Social media is
now not only being used as a content and sharing platform, but also as
a platform for technical support for various of online applications
and services. We have produced a model that can predict server status
based on personality traits and social emotion tones, by investigating
the linear regression and Pearson correlation to identify the key
elements to be used as input for the neural network to build this
model ({\emph{Openness}}, {\emph{Extraversion}},
{\emph{Conscientiousness}}, {\emph{Neuroticism}}, {\emph{Joy}},
{\emph{Sadness}}, {\emph{Anger}} and {\emph{Disgust}}). The model
developed shows a good potential starting point for further data
analysis, with 75\% accuracy in predication based on 57 test cases.

\section{Future Work}\label{futurework}

This paper demonstrates the analysis of one such online application
that has used Facebook as technical support platform for the
users. Such social networks provide substantial textual and
interaction datasets for analysis, providing further insight into
personality traits and social emotion tones.  We are also interested
in profiling complex behaviours and psychopathies using social network
analysis, particularly for crime
informatics~\cite{oatley+crick_fosintsi2014,oatley+crick:2015}. Previous
work in this space analysed the document uploading behaviour (such as
motivation letters, and social media interactions) of the applicants
of the international scholarship mobility portal; by examining the
upload footprint for the users we were able to determine several
classes of behaviour~\citep{oatley-et-al-soccogcomp2015}.

The outcome of the model produced during this work provides the
following recommendations for future research to further incorporate
emotion and personality-based analysis in user-centered modelling, as
well as building upon the availability of high-quality, low-cost and
adaptable tools provided by the IBM Watson Developer Cloud (e.g. IBM
Watson Tone Analyzer and IBM Watson Personality Insights tools),
provide significant further opportunities to integrate linguistic
analysis into this research domain:

\begin{itemize}
\item Expanding the dataset by gathering more data from similar types
  of interactions, as well as technical queries;
\item Annotate and categorise the dataset by gender to investigate the relationship
  between gender and emotion raised by the user in different computer
  system statuses;
\item Explore different computer events not only limited to
  {\emph{Idle}} and {\emph{Failure}}, but including more complex events
      e.g. account hacked, system speed, unexpected error and unsaved data.
\end{itemize}





\bibliographystyle{abbrvnat}
\bibliography{ai2016}

\end{document}